\documentclass[10pt,letterpaper,twocolumn]{article} 

\usepackage{ol2}
\usepackage[draft,implicit=false]{hyperref}
\usepackage{amsmath}

\begin{document}

\twocolumn[ 

\title{\emph{Nonlinearity Induced Critical Coupling}}


\author{K. Nireekshan Reddy$^1$, Achanta Venu Gopal$^2$, and  S. Dutta Gupta$^{*,1}$}

\address{
$^1$School of Physics, University of Hyderabad, Hyderabad-500046, India\\
$^2$DCMPMS, Tata Institute of Fundamental Research, Mumbai-400005, India
 \\
$^*$Corresponding author: sdghyderabad@gmail.com
}

\begin{abstract}
We study a critically coupled system (Opt. Lett., \textbf{32}, 1483 (2007)) with a Kerr-nonlinear spacer layer. Nonlinearity is shown to inhibit null-scattering in a critically coupled system at low powers. However, a system detuned from critical coupling can exhibit near-complete suppression of scattering by means of nonlinearity-induced changes in refractive index. Our studies reveal clearly an important aspect of critical coupling as a delicate balance in both the amplitude and the phase relations, while a nonlinear resonance in dispersive bistability concerns only the phase.
\end{abstract}


 ] 

In recent years there has been a great deal of interest in critical coupling (CC)\cite{cc2,cc3,cc4,cc5,cc6,cc7,cc8} and coherent perfect absorption (CPA) \cite{cpa1,cpa2,cpa3,cpa4}. The physical basis underlying both the above phenomena is near-perfect destructive interference leading to null-scattering. In other words, such systems can only absorb light since both transmission  and reflection  from such structures are suppressed by means of destructive interference. CC uses usually one incident beam with an intricate structure, while CPA uses two beams in simpler configuration to achieve null-scattering. It is clear that such 'perfect' absorption can have applications in varied areas ranging from broadband absorbers to micro-resonator devices \cite{hed,o2}. Recently it was shown how this 'perfect' destructive interference can lead to bending of light the wrong way (on the same side of the normal) \cite{sdgol2012}. There have been fundamental studies on how the absorbed energy is stored in the system at a microscopic level in an opto-mechanical oscillator \cite{gsa}. However, till this day, to the best of our knowledge, there has not been any report on CC in a non-linear layered system, though there are few reports of CC and CPA on other nonlinear systems \cite{*1,*2,o1,o2,o3}. In this letter we address the question of critical coupling in a nonlinear system where one of the constituent layers has a Kerr type nonlinearity. If one thinks of the critical coupling  dip as a resonance, one would expect simple bending of the resonance like in dispersive bistability in nonlinear Fabry-Perot cavities or in nonlinear devices supporting surface plasmons \cite{fp1,fp2,fp4,fp5}.  In this letter we show that such an expectation  does not hold ground because of the rich physics underlying CC or CPA. The dip in the total scattering is due to perfect destructive interference, which depends on a delicate balance of both amplitude and phase. In the context of CPA with a single absorptive slab it amounts to the fact that the interfering waves must have the same amplitude and they must differ by an odd-multiple of $\pi$ phase difference \cite{cpa2}. On the contrary for the case of dispersive bistability, (focusing) nonlinearity leads to an effective increase in the cavity length, resulting in a shift of the resonances to lower frequencies. Thus for a properly detuned system one can have nonlinear response with hysteresis with the possibility of nonlinearity induced modes \cite{mandar}. Note that only Kerr nonlinear phase contributions are essential for the nonlinear response, with no reference to the amplitudes. We show that, CC being a delicate balance involving both the amplitude and the phase, a perfectly tuned linear (at low power levels) system is taken off the critical coupling for increasing input intensities. However, a detuned system can be brought back to critical coupling by means of the nonlinearity induced changes. We demonstrate the above pattern with both the unsplit and split CC resonances \cite{cc4}. Further, for the split resonances only one of the dips is shown to survive at higher powers.
\par
 The structure of the paper is as follows. After a brief survey of the nonlinear characteristic matrix method \cite{fp2}, used to calculate the nonlinear response, we present the results pertaining to the CC resonances for increasing intensities. Our studies clearly reveal the rich physics of the nonlinear critically coupled system and the flexibility it offers for various applications.
\par 
\begin{figure}[htbp]
\center{\includegraphics[width=6cm]{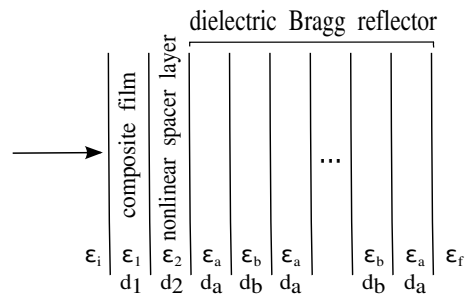}}
\caption{ Schematic view of the layered medium. Background dielectric constant of the nonlinear medium is $\epsilon_2$.}
\label{fig:fig1}
\end{figure}
 We consider the system (shown in Fig.~\ref{fig:fig1})  consisting of a metal-dielectric composite nano layer, a nonlinear spacer layer and a distributed Bragg reflector (DBR). We assume the nonlinearity to be of Kerr type (focusing) with intensity dependent dielectric function \cite{fp2}. A linear counterpart was studied earlier to show CC at single and dual frequencies \cite{cc4}. The DBR, as before, is made up of $2N+1$ layers of dielectric films with alternating dielectric functions and widths. The DBR does not allow any normally incident light with frequency in the rejection band to pass through. The dielectric function $\epsilon_1$ of the metal-dielectric composite film (in the Maxwell-Garnett approximation) is given by \cite{boh}
\begin{equation}\label{eq1}
\epsilon_1\left(\omega\right)=\epsilon_h+\frac{fx\left(\epsilon_m-\epsilon_h\right)}{1+f\left(x-1\right)} , \hspace{0.5cm} x=\frac{3\epsilon_h}{\epsilon_m+2\epsilon_h},
\end{equation}
where $\epsilon_m$ and $\epsilon_h$ are the inclusion and host dielectric constants for the silver-silica composite medium. The parameters for silver are taken from the experimental work of Johnson and Christie \cite{johnson&cristy}. The localized plasmon resonances occur in the rejection band of the DBR and for  certain widths of the spacer layer can lead to $R+T=0$ ($R$ and $T$ intensity reflection and transmission coefficients, respectively). This leads to the usual CC behavior in the linear system heralding near-complete absorption of incident radiation by the composite film.
\par
In oder to calculate the nonlinear response, we make use of the nonlinear characteristic matrix approach \cite{fp2,ray,ray90}, which is valid under the slowly varying envelope approximation (SVEA). In the framework of this theory waves in the nonlinear medium can be expressed as a superposition of the forward and backward propagating waves with amplitude dependent phases \cite{fel}. In particular, the forward and backward waves are characterized by the propagation constants $k_+, k_-$, respectively, expressed as 
\begin{equation}\label{eq2}
k_{\pm}=k_{0}\sqrt{\epsilon_2}\left(1+U_{\pm}+2U_{\mp}\right)^{1/2}, 
\end{equation}
where $k_0=\omega/c$, and $ U_{\pm}=\alpha|A_{\pm}|^2$ are the dimensionless intensities of forward and backward propagating waves, $\alpha$ is the nonlinearity constant and $A_{\pm}$ are the  forward and backward propagating wave amplitudes. As in the standard treatment of optical bistability \cite{fp2,fp4,fp5}, we proceed to calculate the intensities in each layer from the right edge, starting from the transmitted intensity $U_t=\alpha\left|A_t\right|^2$ ($A_t$ is the transmitted wave amplitude). Then $U_{\pm}$ in the nonlinear medium are given by the following equation \cite{ray,ray90}
\begin{equation} \label{eq3}
\begin{pmatrix} 
  \ U_+    \\ 
  \ U_-  
\end{pmatrix}=\left|
{\begin{pmatrix} 
  1     & 1\\ 
  \frac{k_+}{k_0} &  -\frac{k_-}{k_0}
\end{pmatrix}}^{-1}M{_{DBR}}\begin{pmatrix} \ 1 \\ \ \sqrt{\epsilon_f}  \end{pmatrix}\right|^2U_{t},
\end{equation}
 where $M_{DBR}$ is the characteristic matrix for the  DBR \cite{bw}. $\left|~\right|^2$ sign implies element-wise absolute value squared of the vector components. For a given $U_t$ Eq.(\ref{eq3}) can be solved for $U_{\pm}$ by fixed point iteration, and can be used to calculate $k_{\pm}$ (see Eq.(\ref{eq2})). With the knowledge of $k_{\pm}$ one can write all the elements of the nonlinear characteristic matrix \cite{fp2}, the total characteristic matrix is then given by
\begin{equation}\label{eq4}
M=M_{1}\times M_{2}\times M_{DBR},
\end{equation}
 where $M_1$ and $M_2$ are the characteristic matrices for the composite and the nonlinear layers, respectively. In terms of the elements $m_{ij}\left(i, j=1,2\right)$ of the total characteristic matrix $M$ the intensity reflection and the transmission coefficients $R$ and $T$, respectively, are given by \cite{bw,fp1}
\begin{equation}\label{eq5}
R=\left|\frac{\left(m_{11}+m_{12}\sqrt\epsilon_f\right)-\left(m_{12}+m_{22}\sqrt{\epsilon_f}\right)/\sqrt{\epsilon_i}}{\left(m_{11}+m_{12}\sqrt\epsilon_f\right)+\left(m_{21}+m_{22}\sqrt{\epsilon_f}\right)/\sqrt{\epsilon_i}}\right|^2,
\end{equation}
\begin{equation}\label{eq6}
T=\frac{4\left(\sqrt{\epsilon_f/\epsilon_i}\right)^3}{\left|\left(m_{11}+m_{12}\sqrt\epsilon_f\right)+\left(m_{21}+m_{22}\sqrt{\epsilon_f}\right)/\sqrt{\epsilon_i}\right|^2}.
\end{equation}
Calculation of the incident intensity $U_{i}$ is then straight forward
\begin{equation}\label{eq7}
U_{i}=U_{t}/T.
\end{equation}
Eqs. (\ref{eq4})-(\ref{eq7}) can be used to plot total scattering $R+T$ as a function of $U_{i}$ treating $U_t$ as a parameter.
\par
We first discuss some of  the pertinent results of the linear system. For most of our calculations we have used the following parameters, $\epsilon_i=1$, $\epsilon_h=2.25$, $f=0.05$, $d_1=10~nm$,  $\epsilon_2=2.6244$,  $N=10$, $\epsilon_a=5.7121$,  $d_a=42.88~nm$, $\epsilon_b=2.6244$, $d_b=63.72~nm$, $\epsilon_f=2.25$. A detailed study of the linear system revealed that CC resonances occur at $\lambda=407.4~nm$ or at $\lambda=414.1~nm$ for two sequences of $d_2$, namely, $d_2=38.5~nm,  165.0~nm,  291.5~nm \cdots$, and for $d_2=75~nm, 200~nm, 325~nm \cdots$. Note that in each sequence the $d_2$ values have approximately $\lambda/2$ spacing leading to a roundtrip phase shift of about $2\pi$ in the spacer layer. \par
\begin{figure}[t]
\center{\includegraphics[width=8cm]{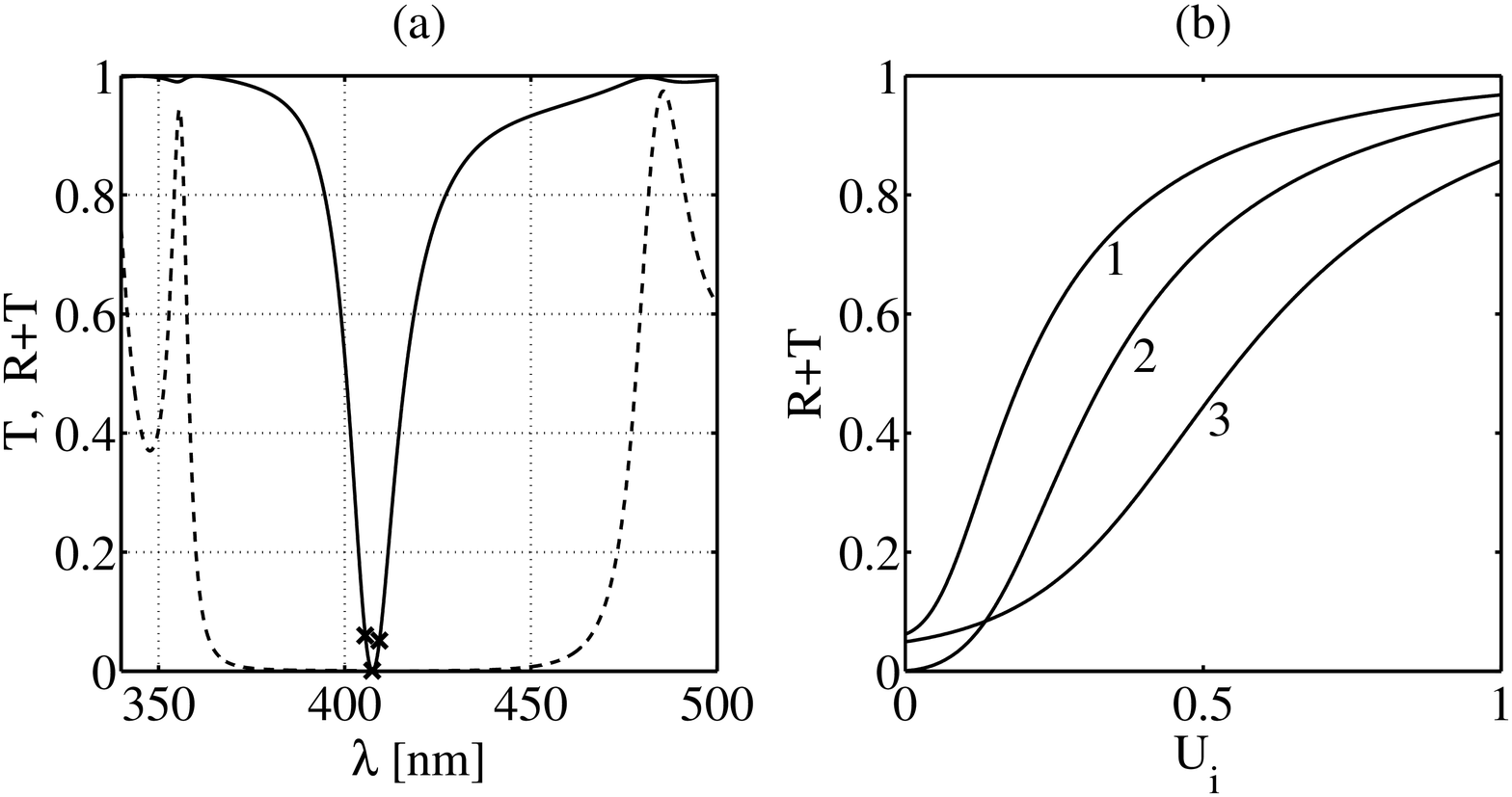}}
\caption{(a) Intensity  transmission coefficient $T$ (dashed curve) and total scattering $R+T$ (solid curve) as functions of the wavelength $\lambda$ for $d_2=75~nm$ for a linear system. (b) $R+T$ as a function of incident intensity $U_{i}$ for three different values of $\lambda$, namely, $405.5~nm$, $407.4~nm$ and $409.3~nm$, marked by crosses in Fig.~\ref{fig:fig2}(a).  Corresponding curves are labeled by 1, 2 and 3, respectively.  Other parameters are as follows, $\epsilon_i=1$, $\epsilon_h=2.25$, $f=0.05$, $d_1=10~nm$, $\epsilon_2=2.6244$, $N=10$, $\epsilon_a=5.7121$, $d_a=42.88~nm$, $\epsilon_b=2.6244$, $d_b=63.72~nm$, $\epsilon_f=2.25$.}
\label{fig:fig2}
\end{figure}
\begin{figure}[t]
\center{\includegraphics[width=8cm]{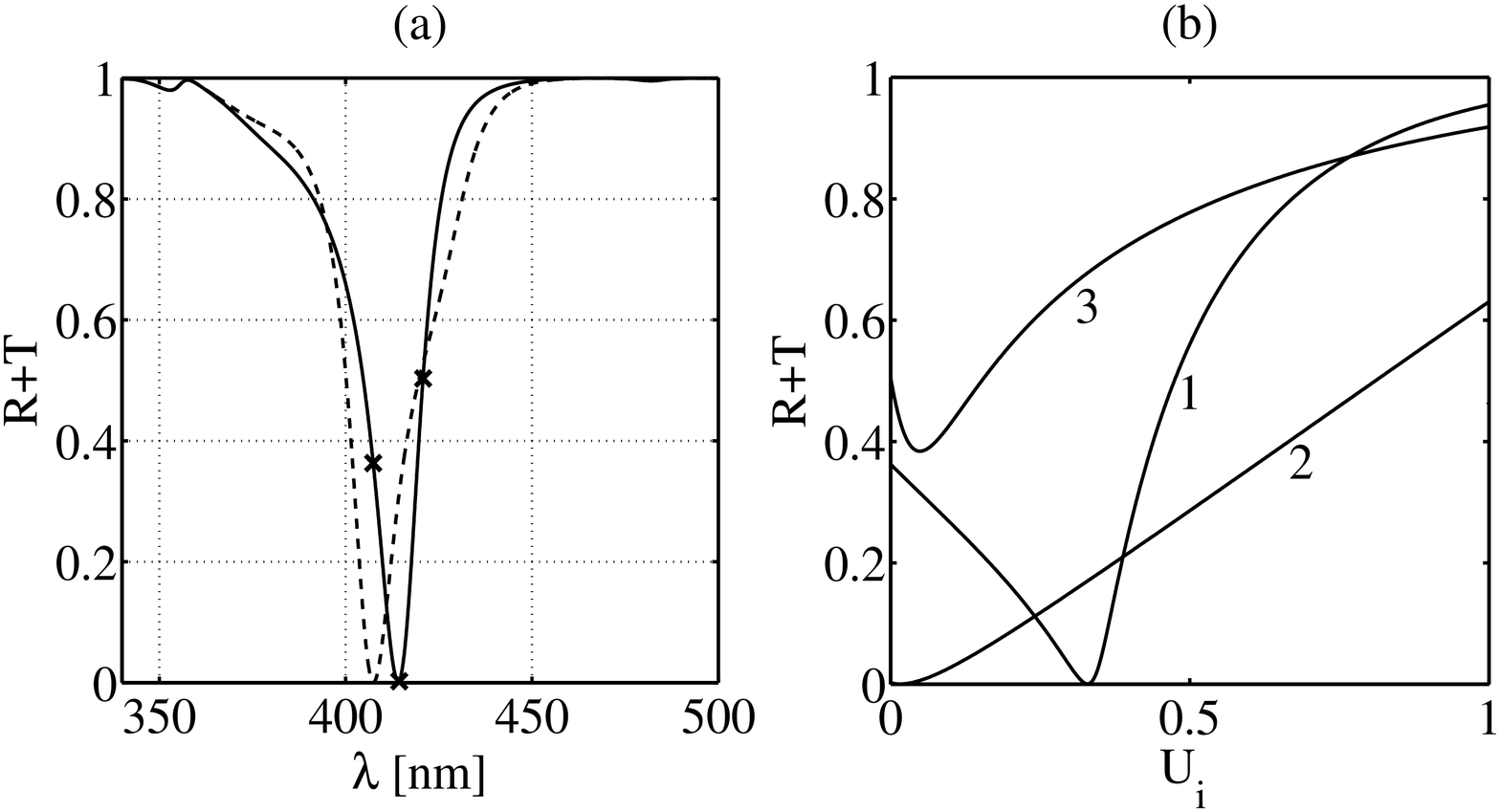}}
\caption{(a) Total scattering $R+T$ as a function of $\lambda	$ for $d_2=165~nm$ (solid curve). (b) $R+T$ as a function of input intensity $U_i$. Curves labeled by 1, 2 and 3, correspond to increasing wavelengths, namely,  $407.4~nm$, $414.1~nm$ and $420.8~nm$, marked by crosses in the Fig.~\ref{fig:fig3}(a). The dashed curve in Fig.~\ref{fig:fig3}(a) is for $U_t=1.6\times 10^{-4}$ which corresponds to the  scattering minimum for curve 1 at $U_i=0.3292$ in Fig.~\ref{fig:fig3}(b). Other parameters are as in Fig.~\ref{fig:fig2}.}
\label{fig:fig3}
\end{figure}
In what follows we present the  result for the nonlinear systems for the two distinct cases, namely, systems tuned at the CC resonance and systems away from the CC resonances. For example, for $d_2=75~nm$ the linear and the nonlinear responses  are shown in Fig.~\ref{fig:fig2}(a) and Fig.~\ref{fig:fig2}(b), respectively.  For the linear response we used the codes for the nonlinear system but at very low power levels. The results were also checked against those of the  code for the linear system. In Fig.~\ref{fig:fig2}(a) we have reproduced the total scattering $R+T$ and the transmission coefficient $T$, which clearly implies a near-total absorption at $\lambda=407.4~nm$. For the nonlinear response we have taken three distinct values of the wavelength, namely, $\lambda=405.5~nm$, $407.4~nm$ and $409.3~nm$, respectively, marked by crosses. For each of these values the effects of nonlinearity induced changes were calculated and they are shown in Fig.~\ref{fig:fig2}(b) by labels 1, 2 and 3, respectively. Note that for each of these values of $\lambda$ (off- or on- minimum), an increase in the input intensity leads to a departure from the critical coupling phenomenon. Thus, in general, nonlinearity inhibits CC in a perfectly tuned linear system. We will show later that there can be departure from this `rule' under specific  conditions. Note that the observed behavior is in sharp contrast with the usual dispersive  bistability resonances, where the near-null dips in reflection or near-total transmission peaks in transmission survive but undergo a distortion (bending) due to nonuniform phase shift at on- and off- resonant wavelengths. The survival of the peaks with near-total transmission is usually observed in nonlinear Fabry-Perot resonances \cite{fp2}. On the contrary, near-zero dips are observed in attenuated total reflection (ATR) studies of nonlinear waveguide or surface plasmon devices \cite{fp5}. Both of these effects owe their origin to the nonlinear phase shift. The situation in CC is more involved, since a delicate balance of both amplitudes and phases is required for CC. Thus any nonlinearity induced change in a perfectly tuned system can upset the conditions for CC leading to substantial scattering from the structure. We could find one exception to this general observation, where the system tuned  at one linear CC resonance at $\lambda=414.1~nm$ could lead to nonlinear CC at the other wavelength $\lambda=407.4~nm$ for specific system parameters. Such a case is shown in Fig.~\ref{fig:fig3}(a) and Fig.\ref{fig:fig3}(b). As before, Fig.~\ref{fig:fig3}(a) shows the response (solid curve) for $R+T$ at low power levels while the curves in Fig.~\ref{fig:fig3}(b) give the intensity dependence. It is clear form Fig.~\ref{fig:fig3}(b) that for $\lambda=407.4~nm$ detuned from the linear CC resonances on one side , one can recover CC in the nonlinear system at a particular intensity level ($U_i=0.3292$) corresponding to $U_t=1.6\times 10^{-4}$. The result for $R+T$ for the $U_t=1.6\times 10^{-4}$ is shown in Fig.~\ref{fig:fig3}(a) with a dashed curve, which shows the resulting nonlinear CC resonance.
\begin{figure}[t]
\center{\includegraphics[width=8cm]{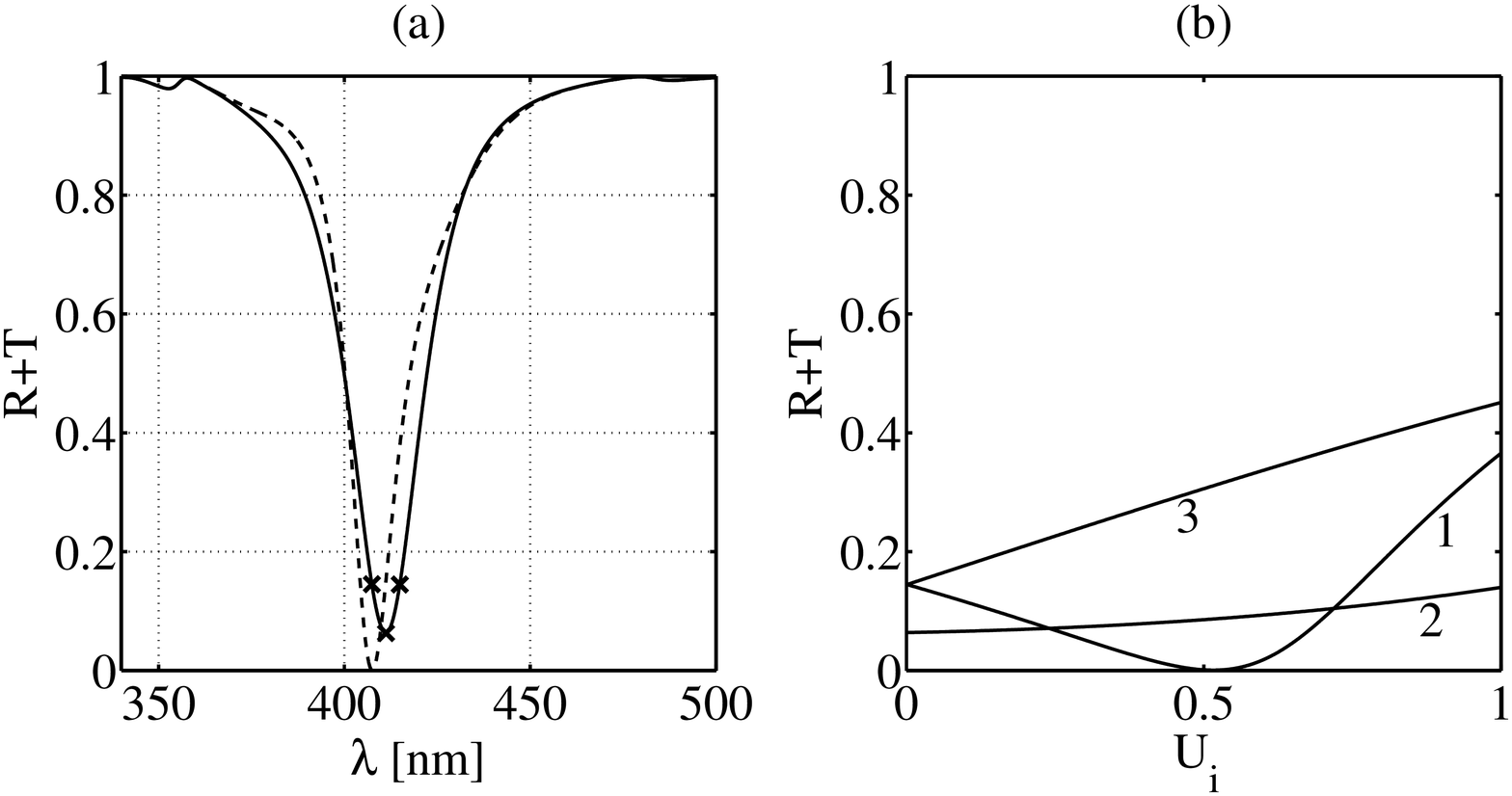}}
\caption{Same as in Fig.~\ref{fig:fig3} except that now $d_2=55~nm$ and the system is away from linear CC resonance. The operating wavelengths for curves labeled by 1, 2 and 3 in Fig.~\ref{fig:fig4}(b) are now $407.4~nm$, $411.2~nm$ and $414.9~nm$, respectively. The dashed curve in Fig.~\ref{fig:fig4}(a) is for $U_t=2.93\times 10^{-4}$ which corresponds to the  scattering minimum for curve 1 at $U_i=0.5164$ in Fig.~\ref{fig:fig4}(b).}
\label{fig:fig4}
\end{figure}
\begin{figure}[t]
\center{\includegraphics[width=8cm]{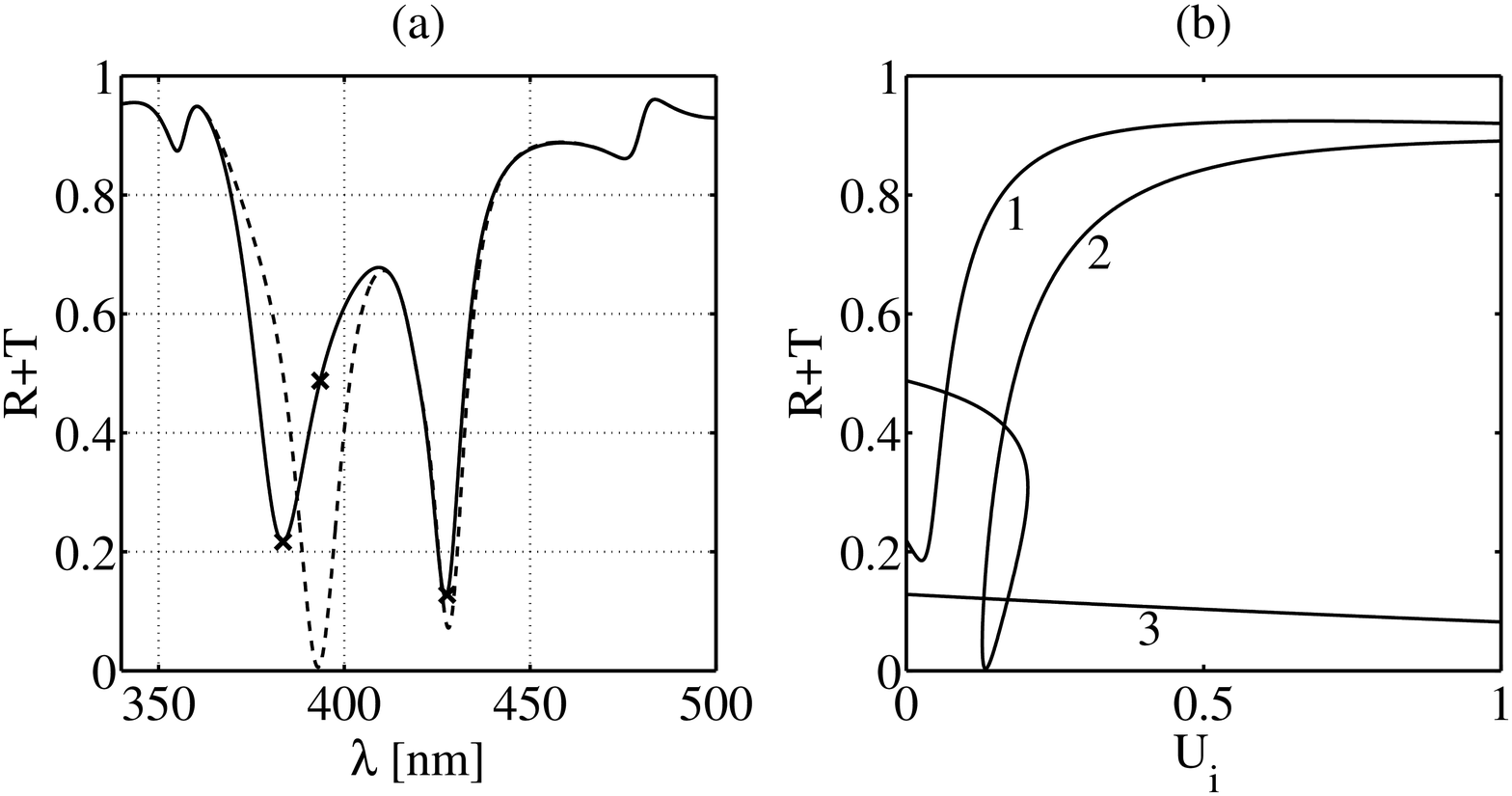}}
\caption{(a) Total scattering $R+T$ as a function of $\lambda	$ for $f=0.1$, $d_1=60~nm$, $d_2=55~nm$ (solid curve). (b) $R+T$ as a function of input intensity $U_i$. Curves labeled by 1, 2 and 3 correspond to wavelengths $384.0~nm$, $393.5~nm$ and $427.7~nm$, respectively (marked by crosses in the Fig.~\ref{fig:fig5}(a)). The dashed curve in Fig.~\ref{fig:fig5}(a) is for $U_t=3.6\times 10^{-4}$ which corresponds to the  scattering minimum for curve 2 at $U_i=0.1333$ in Fig.~\ref{fig:fig5}(b). Other parameters are as in Fig.~\ref{fig:fig2}.} 
\label{fig:fig5}
\end{figure}
\par
We now concentrate on a system with parameters detuned from a linear CC resonance (see Fig.
~\ref{fig:fig4}(a)) and show how nonlinearity induced changes can tune the system back to CC (see Fig.~\ref{fig:fig4}(b)). The null scattering point in curve 2 in Fig.~\ref{fig:fig4}(b) corresponds to $U_i=0.5164$, $U_t=2.93\times 10^{-4}$. The dashed curve Fig.~\ref{fig:fig4}(a) confirms the observed nonlinear CC phenomenon with $U_t=2.93\times 10^{-4}$. 
\par
Further we turn our attention to the CC at dual frequencies, reported earlier \cite{cc4}. We show that recovering CC at both the frequencies may not be an easy task, while the system can be tuned at one CC frequency by increasing the incident wave intensity. This is shown in Fig.~\ref{fig:fig5}. It is evident from Fig.~\ref{fig:fig5} that a system tuned at the minima of $R+T$ may not lead to nonlinear CC phenomenon, while a system tuned at another frequency, for example $\lambda=393.5~nm$, can lead to CC with interesting hysteresis effects. The dashed curve in Fig.~\ref{fig:fig5}(a) again confirms the complete suppression of scattering at $\lambda=393.5~nm$ for $U_i=0.1333$ ($U_t=3.6\times10^{-4}$).
\begin{figure}[t]
\center\includegraphics[width=6cm]{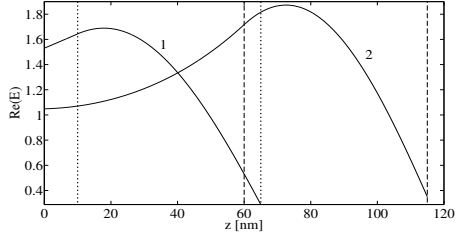}
\caption{$Re(E)$ as a function of $z$ in the composite and the nonlinear layers for two cases of nonlinearity-induced CC, namely, (1) minimum of curve 1 in Fig.~\ref{fig:fig4}(b) with $f=0.05$, $d_1=10~nm$, $d_2=55~nm$, $U_i=0.5164$ ($U_t=2.93\times 10^{-4}$), (2) minimum of curve 2 in Fig.~\ref{fig:fig5}(b) $f=0.1$, $d_1=60~nm$, $d_2=55~nm$, $U_i=0.1333$ ($U_t=3.6\times 10^{-4}$). Vertical lines demarcate the interfaces, dotted (dashed) line for case 1 (case 2). Other parameters are as in Fig.~\ref{fig:fig2}.} 
\label{fig:fig6}
\end{figure}
\par
We also looked at the field distributions (shown in Fig.~\ref{fig:fig6}) for two on-resonant cases  corresponding to the CC minimum in Fig.~\ref{fig:fig4}(b) for curve 1 and in Fig.~\ref{fig:fig5}(b) for curve 2, respectively. It is clear that the field distributions mimic those for the linear systems\cite{cc4}, depicting the build up of the field due to nonlinear CC resonances.
\par
In conclusion, we have studied critical coupling in a nonlinear layered medium consisting of a metal-dielectric composite nano layer, a Kerr-nonlinear spacer layer and a distributed Bragg reflector. The linear counterpart was studied earlier to show complete suppression of scattering from the system at specific wavelengths. Our studies reveal that the nonlinear structure offers rich physics and novel flexibilities in controlling the CC resonances. In particular, nonlinearity-induced changes can restore CC resonance in a linear off-resonant system. On the contrary, a linear critically coupled system is usually taken off-resonance by nonlinearity with one exception when linear CC at one spectific wavelength (system dependent) can evolve into a nonlinear CC at another wavelength. Possibility of hysteretic response is also demonstrated. The above results are corroborated with calculations of intensity dependent response based on a nonlinear characteristic matrix formalism valid under SVEA. Departure of the system from CC for higher intensities was explained in terms of the delicate balance of both amplitude and phase involved in any CC resonance. In contrast, the standard nonlinear resonances in dispersive bistability concerns only the phase relations and nonlinearity `bends' the resonance retaining the transmission peaks in cavities or the reflection dips in ATR studies. Our results can find varied applications in nano- and micro- devices for switching, sensing, etc.
\par
One of the authors (SDG) would like to thank Girish Agarwal for discussions.

\end{document}